\documentclass{aa}
\usepackage{graphicx}

\usepackage{txfonts}
\usepackage{xcolor}
\newcommand{\norm}[1]{\left\lVert#1\right\rVert}

\newcommand{\onenorm}[1]{\norm{#1}_1}

\newcommand{\nuclearnorm}[1]{\norm{#1}_{*}}
\newcommand{\frobeniusnorm}[1]{\norm{#1}_{F}}
\usepackage{geometry}
\usepackage{array}
\usepackage{longtable}
\usepackage[colorlinks=true,urlcolor=blue,citecolor=blue,linkcolor=blue]{hyperref}
\usepackage{graphicx}
\usepackage{array}
\usepackage{color}
\usepackage{colortbl}
\usepackage{subcaption}
\usepackage{multirow}
\usepackage{placeins}
\usepackage{natbib}
\bibpunct{(}{)}{;}{a}{}{,} % to follow the A&A style

 \usepackage{wasysym}
\begin{document} 

   \title{Thermal background reduction for mid-infrared imaging by\\ low-rank background and sparse point-source modelling}

   \subtitle{Quantitative and qualitative demonstrations on\\ ground-based VISIR and airborne SOFIA time series}

   \author{R.A.R. Moens\inst{1}
          \and A.G.M. Pietrow\inst{2,3,4}
          \and B. Brandl\inst{4,5}
          \and R. Van de Plas\inst{1,6,7}
          }

   \institute{Delft Center for Systems and Control, Delft University of Technology,
              Mekelweg 2, 2628 CD Delft, Netherlands\\
              \email{r.a.r.moens@tudelft.nl / rogermoens@delfius.nl}
         \and
            Leibniz-Institut für Astrophysik Potsdam (AIP), An der Sternwarte 16, 14482 Potsdam, Germany.
        \and
            Centre for Mathematical Plasma Astrophysics, KU Leuven, Celestijnenlaan 200B, B-3001 Leuven, Belgium.
        \and
            Leiden Observatory, Leiden University, PO Box 9513, 2300 RA Leiden, Netherlands.
        \and
            Faculty of Aerospace Engineering, Delft University of Technology, Kluyverweg 1, 2629 HS Delft, Netherlands.
        \and
            Mass Spectrometry Research Center, Vanderbilt University, 465 21st Ave S. Room 9160, Nashville, TN 37232, USA.
        \and Department of Biochemistry, Vanderbilt     University, 465 21st Ave S. Room 9160,      Nashville, TN 37232, USA.
        }

   \date{  }
 
  \abstract
   {Mid-infrared astronomy from the ground faces critical challenges in accurately detecting and quantifying sources due to the dominant spatially and time-variable background noise. Moreover, chopping and nodding, the traditional methods for dealing with these background issues, will not be technically feasible on the next generation of extremely large telescopes. This limitation requires the development of novel computational methods for a robust background reduction.}
   {We present and evaluate a novel method named LOw-RAnk Background ELimination (LORABEL) to improve the sensitivity of mid-infrared astronomical observations, without the need for classical telescope nodding, source masking, or other overheads in observing time.}
   {We applied a low-rank background-reduction strategy to (1) data taken on the ground with the VLT Imager and Spectrometer for mid-InfraRed (VISIR) with synthetically injected sources, and (2) airborne data from the Stratospheric Observatory for Infrared Astronomy (SOFIA). We compared the performance of our new method to classical chopping and nodding techniques, and analysed the effect on source photometry and detection precision for different observational scenarios.}
   {In regimes with a low signal-to-noise ratio (S/N $<5$) in the ground-based VISIR data, LORABEL reduces variation in the photometric error with respect to chopping differences alone and even the classical chop-nod sequence, at the cost of introducing a bias. Secondly, we demonstrate that LORABEL increases detection precision in comparison to traditional background-reduction methods. For the SOFIA dataset, we achieve a $20-100$ fold decrease in mean background flux with respect to the traditional chop-nod method while preserving most of the source flux. Our findings suggest that LORABEL is applicable to a wider range of instrumental observation, that is, both ground-based and airborne, and it is a suitable tool in the context of faint-source detection.}
  {}


   \maketitle
   
\section{Introduction}
The thermal mid-infrared (mid-IR; $\lambda \approx 3$ -- $40~\mu$m) covers an astronomically unique part of the electromagnetic spectrum that is ideal for studies of warm dust, molecules in the interstellar medium, heavily obscured objects (protostars and active galactic nuclei), temperate exoplanets, Solar System objects, and redshifted objects at large distances. Key questions addressed by mid-IR instruments include the evolution of star-forming regions, the compositions and surface temperatures of various celestial objects, solar flare electron densities, and the effect of molecular clouds on star formation and galactic dynamics.
Historically, mid-IR imaging systems have evolved from Earth-based and airborne to space-based platforms, with notable instruments including the VLT Imager and Spectrometer for mid-Infrared (VISIR; \citealt{rio1998visir}), the Texas Echelon Cross Echelle Spectrograph (TEXES; \citealt{lacy2002texes}), CanariCam (\citealt{Telesco2003}), the Stratospheric Observatory for Infrared Astronomy (SOFIA; \citealt{krabbe2000sofia}), the Infrared Astronomical Satellite (IRAS; \citealt{irace1983iras}), and the Spitzer Space Telescope (\citealt{werner2004spitzer}). The most impressive images at mid-IR wavelengths have recently been delivered by the James Webb Space Telescope, \citep[JWST,][]{rieke2015mid} with its 6.5m aperture. While the boundary conditions for observing at mid-IR wavelength with a cooled telescope in space are unsurpassed in terms of thermal background and stability, the mission costs are usually extremely high, and the telescope aperture size (key to achieving high angular resolution) has to be much smaller than a comparable facility on the ground. Very high angular resolution, which, for instance, is essential to directly image Earth-like exoplanets, requires a telescope aperture much larger than 10 meters. Furthermore, ground-based and airborne instruments are easily accessible, which allows us to modify and extend operational lifespans at relatively low cost.
However, warm ground-based telescopes face additional challenges, most notably the large thermal background, which can dominate the source signal by a large factor and complicates the source detection \citep{burtscher2020towards}. In the best case, observational strategies and subsequent calibration measures permit the subtraction of the average thermal background, leaving only the Poisson-distributed photon shot noise. However, the complexity and large number of variable thermal emitters in (or close to) the telescope beam have often led to residual background patterns in addition to the photon shot noise floor, which significantly reduced the detection limit. This leads to an unfortunate situation: while the new facilities on the ground will provide the angular resolution to directly image exoplanets near their host stars, insufficient background subtraction could lead to insufficient sensitivity to detect these exoplanets. Therefore, the elimination of residual background patterns is crucial for the detection of exoplanets and many other faint sources of interest \citep{Petit2020,wagner2021imaging, matthews2024temperate}.
The origin and exact timescales of these thermal background patterns have many different causes, are often even unknown, and differ from one telescope to the next. This makes their elimination difficult and sometimes impossible. Generally, the thermal background can be categorised into (1) highly time-variable (sub-second scale) photon shot noise, (2) slowly variable (second scale) background emission patterns of low spatial frequencies, and (3) underlying quasi-static (minute scale) patterns. To first order, the origin of (1) is the thermal (grey-body) emission from the atmosphere and the warm telescope, amplified by variations in detector responsivity, (2) is due to changes in telescope configuration (e.g. due to tracking) and environmental conditions, and (3) is due to the optical configuration (e.g. baffles). Traditionally, the background is reduced through the so-called chopping and nodding technique \citep{Papoular1983}. However, this method is not always feasible, especially not on the next generation of telescopes, like ESO's Extremely Large Telescope (ELT). There, chopping is impossible due to the sheer size of the secondary mirror, and classical nodding is not feasible because ELT is not a stiff structure when tracking the object on the sky, but rather an actively controlled, dynamic five-mirror-telescope that is constantly being realigned. Hence, new instruments, like the Mid-infrared ELT Imager and Spectrograph \citep[METIS, ][]{brandl2012metis}, need alternative background-reduction techniques. METIS will use a compact and fast internal chopper to take care of the fast variations (1), but requires additional methods to account for the slower variations, which have traditionally required nodding. 

 Recent work \citep{Pietrow2016, pietrow2019inverse} on slightly more sophisticated chopping sequences has shown that novel reduction schemes can provide good results with chopping alone and no additional telescope nodding. Other techniques are based on computationally modelling the chop residuals by \textcolor{black}{principal component analysis (PCA) \citep{rousseau2024improving}} or by drift scanning \citep{2014Heikamp,ohsawa2018slow,torres2021canaricam}. However, these methods only partially take the available information into account, require knowledge of {the position of the field of view}, timescales of background variation, and drift parameters, or decrease the observational efficiency altogether. 

 To address these issues, we propose an algorithm that splits the data into a low-rank background image, a sparse signal, and a random noise component, similar to a method used in high-contrast imaging \citep{gonzalez2016low}. The method presented here uses chop-only measurements of point sources. The advantage of our approach is that it leverages the mathematical properties of the available data without requiring specific knowledge of point-source positions, the background patterns, or their variability timescales, for example. In addition, the method we developed provides a path forward for modelling and reducing residual background patterns due to yet unknown boundary conditions for instance for METIS at the ELT, where background structures, which cannot be reduced by classical methods, may occur \citep{sauter2024detection}\par

We evaluate the photometric performance of the method we propose for point sources by (1) comparing it to artificially injected point sources in empty ground-based VISIR data and (2) comparing it to chop-only and chop-nod methods in airborne SOFIA data to show its versatility. The structure of the paper is as follows: we first introduce the definition of terms used throughout this paper, followed by the introduction of the observational data and the method, including the specifics of aperture photometry. Subsequently, we present the results for both case studies, and finally, we summarise and discuss our findings in the conclusions.

%--------------------------------------------------------------------
\section{Definitions}
\subsection{Nomenclature}
Throughout our analysis, we use a number of specific terms.  The following glossary provides brief definitions of these terms in the context of this paper.
\begin{itemize}
    \item Background signal: any undesired signal. This can be of astronomical (e.g. clouds in front of sources, but also satellite flybys and perturbations caused by the Earth's atmosphere), instrumental (e.g. interaction of the signal with instrument components or telescope) and computational nature (e.g. introduced by the data-reduction pipeline).
    \item Chop-nod scheme: scheme used to perform subtractions of measurements or time frames in different chop and nod positions to obtain an optimal background reduction.
    \item Chop-nod subtraction: subtraction obtained by subtracting two chop subtractions obtained from nodding positions.
    \item Chop-only regime: regime under which only chopping but no nodding is possible.
    \item Chop position: off-axis tilt position of the secondary mirror. We use chop position in this paper to denote the secondary mirror position of a measurement or time frame.
    \item Chop subtraction: subtraction of two time frames in different chop positions.
    \item Chopping: performing a secondary mirror tilt to record off-axis measurements. Due to the off-axis tilt, sources move relative to the time frame. Chopping is often performed at a frequency of 1 Hertz or more.
    \item Flux: average number of photons passing through a unit area per unit time.
    \item Flux density: the density of flux (or photons) over a given area, measured in units of photons per square meter per second, or analog-to-digital units (ADUs) per square arcsecond.
    \item Co-addition: the process of combining multiple measurements over time to improve the signal-to-noise ratio and reduce random variations in the data.
    \item Measurement: a non-integrated or non-added recording of a particular field of view of interest, in our case, a 2D image, measuring flux.
    \item Nod position: position of the primary mirror. We use nod position in this paper to denote the primary mirror position of a measurement or time frame.
    \item Nodding: performing a primary mirror shift. Due to this shift, sources move relatively to the time frame. Nodding is often performed on a timescale of $1$ to $2$ minutes.
    \item Point source: a source without spatial structure, whose morphology is simple, given by the optical performance of the imaging system. In this paper, we solely deal with point sources. Ideally, they appear as airy patterns in the measurements or time frames. We consider them to be small with respect to the field of view. Perfect background-subtracted measurements, or time frames, can thus be considered to be sparse.
    \item Source signal: signal of interest. In this paper, we only deal with point sources as signals of interest.
    \item Pixel: we consider a pixel a single element/entry of an individual time frame.
    \item Time frame: an individual time frame is the result of the co-addition of measurements, for example to counteract noise variations.
    \item Time-frame average: average of a time series, that is, representing flux densities across a 2D image.
    \item Time series: a set of time frames recorded over a particular time span. Time series consist of different time frames, one for each point in time.
\end{itemize}

\subsection{Mathematical notations}\label{maths_notations}
We denote a scalar with a lowercase letter $a \in \mathbb{R}$, a vector of size $m$ with a bold lowercase letter $\mathbf{a} \in \mathbb{R}^m$, and a matrix of size $m$-by-$n$ with a capital letter $A \in \mathbb{R}^{m\times n}$. The vectorising operation $\text{vec}(A)$ acts on a matrix and transforms it into a vector by stacking all columns or rows into a single column or row: $\mathbb{R}^{m\times n} \rightarrow \mathbb{R}^{mn}$. Furthermore, we use $||A||_*$ to denote the nuclear norm of a matrix $A$, $||A||_F$ for the Frobenius norm,  {$||\mathbf{a}||_1$ as vector 1-norm}, and $||\mathbf{a}||_2$ as vector 2-norm. The absolute value is denoted as $\left|\cdot\right|$. The rank of a matrix $A$, $\text{rank}(A)$, denotes the dimension of the subspace containing the matrix $A$. Generally, we talk about low-rank when $\text{rank}(A) \ll \min{(m, n)}$. The low-rank property of matrices has several favourable mathematical properties and can be used to describe data from the smallest possible set of bases. The sparsity of a matrix is considered to be the number of non-zero values in its entries. High sparsity implies many zero values. When we talk about a sparse regime, this implies that a matrix or vector contains many zero values in their entries. When we consider a small dense noise matrix $A$, we imply that  $||A||_F \leq \delta$, that is, small in Frobenius norm, and that it has non-zero values in (almost) all entries (low sparsity), that is, it is dense. Finally, we use superscript letters $\mathbf{a}^j$ to denote chop positions, subscripts letters $\mathbf{a}_t$ to denote time frames, and bars ($\bar{\mathbf{a}}$) and hats ($\hat{\mathbf{a}}$) for different nod positions. Furthermore, when we consider a dense noise matrix, the noise is considered to be independent and identically distributed (i.i.d.).

\subsection{Noise and all of its friends}
Mid-infrared astronomical observations are affected by two primary noise categories: systematic and {statistical} noise, both considered background signal in this paper. Systematic noise arises from instrumental artefacts, detector imperfection, such as detector response (flat-field) variations, clusters of dead pixels, channel offsets, or non-linearities, and differences in the illumination of the optical components due to the beam offsets between the different chopping positions, creating pseudo-consistent structured variations across measurements that primarily depend on local conditions and telescope orientation. Statistical noise, on the other hand, follows a Poisson detection process and is present in sources as well as in background. We model this {statistical} noise for the background using an i.i.d. Gaussian approximation. In this framework, i.i.d. means that each noise sample is statistically independent and drawn from the same Gaussian probability distribution. While Poisson noise fundamentally represents discrete integer-based photon counting with variance equal to the mean, the Gaussian approximation provides a continuous computationally tractable representation for signal reconstruction, which is particularly valid when photon counts are sufficiently high. Here, statistical noise denotes the aggregate of photon shot noise from all incident flux (source, background, and internal radiation) together with detector read-out and dark-current noise, which we approximate as the square root of the total counts.

\section{Observational data, source detection, photometry, and metrics}

\subsection{VISIR data case study}
The VLT Imager and Spectrometer for mid-InfraRed \citep[VISIR, ][]{rio1998visir} is a ground-based mid-infrared observational instrument mounted to the Cassegrain focus of the second unit telescope of the Very Large Telescope (VLT). This telescope is one of four telescopes with a diameter of 8.2 meters that make up the core of the Paranal Observatory in Chile. VISIR provides high-resolution imaging and spectroscopy in the mid-IR wavelength range \citep{rio1998visir}. Our {raw} VISIR dataset was acquired as technical time observations \citep{Pietrow2016}. The dataset used in our first case study, {referred} to here as empty VISIR data, only contains background and no source signal, as it was taken pointing close to zenith without any target. The objective was to recreate an idealised dataset based on raw VISIR data to study our method under idealised circumstances. 
As an instrument mounted at the Cassegrain focus, the image will rotate, but the pupil will stay fixed. If the derotator is turned on, the pupil will rotate accordingly to compensate for image rotation. The latter setting is sub-optimal (and thus, not recommended) for the general use of the presented algorithms, unless the integration times are short and the corresponding pupil rotation is quite small. However, for the analysis of the background emission we present here, telescope tracking and image derotation were both turned off, and the pupil configuration was constant.

These {raw VISIR data were therefore firstly chop-subtracted} and secondly low-rank approximated by performing a rank-$7$ singular value decomposition (SVD), defining a low-rank ground-truth background. Secondly, we added Gaussian i.i.d. noise to our measurements with $\mu=0$ and $\sigma=1$ to mimic Poisson-noise-related effects. The low rank and the injection of Gaussian i.i.d. noise allowed us to optimally set the method parameters (see method). Finally, we injected sources (see the specifics below and in Table \ref{table:instrument_specs}).

\begin{table*}[h!]
    \caption{\label{table:instrument_specs} VLT/VISIR empty field data and SOFIA/FORCAST $\gamma$ Cygni observations.}
    \centering
    \renewcommand{\arraystretch}{1.5}
    \begin{tabular}{|>{\raggedright\arraybackslash}p{4cm}|>{\raggedright\arraybackslash}p{6cm}|>{\raggedright\arraybackslash}p{6cm}|}
    \hline
    \rowcolor[gray]{0.9} Specifications & VLT/VISIR & SOFIA/FORCAST \\
    \hline
    \hline
    \rowcolor[gray]{0.9} Astronomical Specifics & & \\
    \hline
    Object & Injected point source of varying intensity & $\gamma$ Cygni\\
    \hline
    Pointing & RA 276.743637 deg, DEC -25.43501 deg & RA 330.472792 deg, DEC 48.731228 deg \\
    \hline
    Time of Recording & 2016-03-22T23:37:35.7488 & 2022-05-11T09:22:10.584\\
    \hline
    Observation Time & $\sim 43$s & $\sim 14$min \\
    \hline
    \hline
    \rowcolor[gray]{0.9} Instrumental Specifics & & \\
    \hline
    Chop Directions & 0, 90, 180, and 270 deg & 0, 90, 180, and 270 deg \\
    \hline
    Chop Frequencies & 4 Hz & 1, 4, and 5 Hz \\
    \hline
    Nod Directions & Up-down and left-right configuration & Up-down and left-right configuration \\
    \hline
    Spectral Filter(s) & J8.9 & F088, F111, F197 \\
    \hline
    \hline
    \rowcolor[gray]{0.9} Preprocessing Specifics & & \\
    \hline
    Co-addition & $0.0125$s integration time, 7 sub-integrations & 48.4406 sampled measurement rate, 22 co-additions in original data production \\ 
    \hline
    Other Preprocessing & None & Flat-fielding, dark current subtraction, bad pixel correction, data is normalized for co-adding \\
    \hline
    \hline
    \rowcolor[gray]{0.9} Mathematical Specifics & & \\
    \hline
    Data Sizes & $900\times1024$ pixels for 171 time frames, i.e. a data matrix $A \in \mathbb{R}^{921,600 \times 171}$ & $256\times256$ pixels for 24 time frames for 2 nod positions and 2 chop directions, i.e. 4 data matrices $A_i \in \mathbb{R}^{65536 \times 24} \ \forall i \in [1,4]$ \\
    \hline 
    \end{tabular}
    \tablefoot{
The table provides a comprehensive overview of the astronomical, instrumental, pre-processing, and mathematical specifics for both datasets. Astronomical specifics include target coordinates and observation times. Instrumental details describe the chop/nod configurations, frequencies, and spectral filters we used. The pre-processing section covers integration times, co-addition, flat-fielding, and bad-pixel correction, ensuring optimised data quality. Mathematical specifics detail the data matrix structures used for analysis, including spatial and temporal specifics of the recorded data. Background variations for VLT/VISIR over a timescale of 43s are not representative of a typical 30 min integration.
}
\end{table*}

\subsection*{Injected sources}
The injected sources were modelled by two $13$-pixel full width at half maximum (FWHM) Gaussian profiles, mimicking positive and negative sources of chopped VISIR images. To test our methods under the most challenging conditions, we adopted a Q-band example with a point-source FWHM of 13 pixels, which is larger than typical N-band VISIR observations. The scale of these Gaussian profiles was derived by calculating the FWHM as
\begin{equation}
\label{eq:FWHM}
\begin{aligned}
\text{FWHM} \approx 1.22 \frac{\lambda}{D},
\end{aligned}
\end{equation}
where $D$ is $8.2\ $m for VISIR, and $\lambda = 20\ \mu$m was defined to be in the Q band. This led to an FWHM of approximately $2.98\times 10^{-6}$ radians, or around $0.6$ arcsec. With the VISIR pixel scale being $0.045$ arcsec/pixel in the small-field mode  \citep[SFM, ][]{esoOverview}, this leads to a $13$-pixel FWHM. The injected sources were normalised and then multiplied by the desired total source flux, $F_{\text{source}}$.

\begin{figure}[htbp]
  \centering
  \includegraphics[width=.8\linewidth,keepaspectratio]{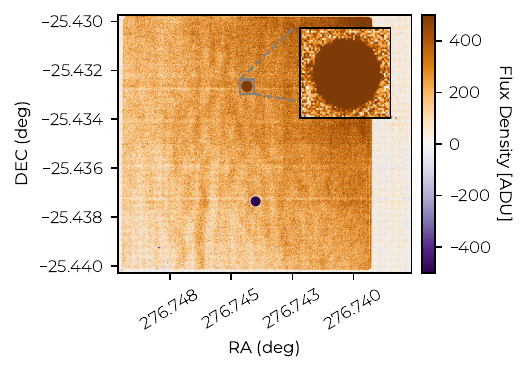}
  \caption{Time-frame average for the VLT/VISIR case study with injected sources. {The display scale has been deliberately compressed to emphasize background structures. As a result, the positive source appears saturated and does not show a distinct peak} (see Table \ref{table:instrument_specs} for the data specifics).}
  \label{fig:visir_overview_image}
\end{figure}

{This source flux was obtained by multiplying the square root of the number of pixels used in the aperture photometry ($n_{\text{pix}}$), 
the standard deviation $\sigma_{\text{bkg}}$ of a square aperture at the position of injection in the chop residual, and the injection signal-to-noise ratio (S/R) as follows:}
\begin{equation}
F_{\text{source}} = \text{S/R}_\text{injection} \times \sqrt{n_{\text{pix}} \cdot \sigma_{\text{bkg}}^2}.
\end{equation}
The location of the source injection was randomly generated for each experiment, and the positive and negative sources were generated minimally $3\times \text{FWHM}+30$ pixels apart. The size of the background square at each injection position (for which the standard deviation was calculated) was defined as $3\times \text{FWHM}+15$ pixels. To yield a variety of source strengths for evaluating our methods, the $\text{S/R}_\text{injection}$ took the following values: $1$, $2$, $3$, $5$, $10$, $20$, and $50$ for evaluating the photometric error and values $1$ to $10$ for detection evaluation. Importantly, we used the chop residual as reference to calculate the standard background deviation. An example of an injection in the time-frame average is given in Fig. \ref{fig:visir_overview_image}

\subsection{SOFIA data case study}
\begin{figure}[htbp]
  \centering
  \includegraphics[width=.8\linewidth,keepaspectratio]{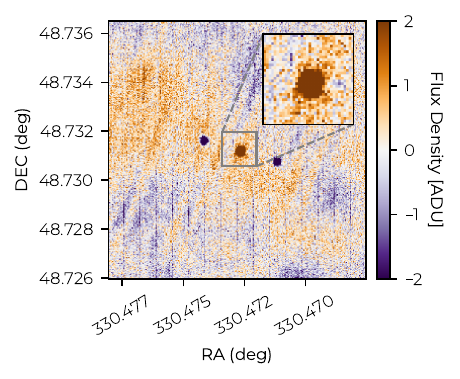}
  \caption{Time-frame average for the SOFIA/FORCAST case study (see Table \ref{table:instrument_specs} for the data specifics). {The second negative source appears because the data were acquired using the inverse-chopped technique \citep{pietrow2019inverse}.}}
  \label{fig:sofia_overview_image}
\end{figure}

The Stratospheric Observatory for Infrared Astronomy \citep[SOFIA,][]{krabbe2000sofia} was an airborne $\varnothing$2.5~m telescope inside a modified Boeing 747SP, which was active from 2010 to 2022. One of its instruments was the Faint Object infraRed CAmera for the SOFIA Telescope \citep[FORCAST, ][]{Herter2018forcast},  a dual-channel mid-infrared camera and spectrograph sensitive from \mbox{5~--~40~$\rm\mu$m}. SOFIA could operate at altitudes of up to $14$ km, above most of the water vapour in Earth's atmosphere, thereby providing clear views of the infrared sky. The observations taken under the mission ID \texttt{2022-05-11\_FO\_F867} and plan ID \texttt{71\_0025} were obtained during technical time and observed $\gamma$ Cygni. The specifics of the SOFIA dataset are also given in Table \ref{table:instrument_specs}. While SOFIA still requires background subtraction similar to ground-based facilities, the thermal background is significantly reduced due to the lower airmass above the aircraft. The objective is to show the performance of our method in non-ideal circumstances and to show its versatility (not only restricted to ground-based observations{, but also airborne observations}). The time-frame average is given in Fig. \ref{fig:sofia_overview_image}

\subsection{Source detection}\label{detection}
{For the sake of clarity, source detection is introduced here as it is a necessary step for large-scale automatic testing and bootstrapping.} In this paper, source detection was carried out using the DAOStarFinder algorithm, provided by the Python package Photutils \citep{bradley2016photutils}. The DAOStarFinder parameters considered were (1) the FWHM, and (2) the detection threshold. The FWHM parameter was fixed at $13$ pixels for the chop-only and chop-nod methods, whereas it was set to $11$ pixels for LORABEL. The threshold, denoted by $\epsilon$, was computed using the following relation:
\begin{equation}
\label{eq:threshold}
\epsilon = f \times \sigma_c \times \text{S/R}_\text{injection},
\end{equation}
where $f$ is a factor used to tune the detection sensitivity, $\sigma_c$ is the standard deviation derived from a $3\sigma$ clipping procedure of the whole time-frame average, and $\text{S/R}_\text{injection}$ is the resulting signal-to-noise ratio of the injected source. The values of the threshold factor $f$ used in this study are summarised in Table \ref{tab:threshold_factors}. This threshold factor and threshold affect the outcomes of the DAOStarFinder algorithm. As we injected sources using a local standard deviation measure, this might sometimes conflict with the global standard deviation, which is defined over the whole field of view and is used in Eq. \ref{eq:threshold}, because this lowering of the threshold is often necessary to obtain better results. Finally, we filtered the results obtained from DAOStarFinder to reduce false positives by requiring roundness1 and roundness2 to be $<0.3$ and the peak S/R of the detected object $>2$.

\begin{table}[htbp]
\renewcommand{\arraystretch}{1.2}
\caption{\label{tab:threshold_factors}Threshold factor ($f$) values.}
\centering
{\small
\begin{tabular}{|c|c|c|c|}
\hline
\rowcolor[gray]{0.9} S/R$_{\text{injection}}$ & Chop-Only & Chop-Nod & LORABEL \\
\hline \hline
$1$  & $10^{-10}$ & $10^{-10}$ & $10^{-10}$ \\
\hline
$2$  & $10^{-10}$ & $10^{-10}$ & $10^{-10}$ \\
\hline
$3$  & $10^{-10}$ & $10^{-10}$ & $10^{-10}$ \\
\hline
$4$  & $0.6$ & $0.8$ & $0.7$ \\
\hline
$5$  & $0.6$ & $0.8$ & $0.7$ \\
\hline
$6$  & $0.5$ & $0.7$ & $0.6$ \\
\hline
$7$  & $0.4$ & $0.7$ & $0.5$ \\
\hline
$8$  & $0.4$ & $0.7$ & $0.5$ \\
\hline
$9$  & $0.4$ & $0.7$ & $0.5$ \\
\hline
$10$ & $0.4$ & $0.7$ & $0.5$ \\
\hline
\end{tabular}
}
\tablefoot{Values used for different detection methods as a function of the injected source S/R. The very low values, i.e. $10^{-10}$, at low injection S/Rs reflect the idea of at least detecting the true-positive source, where at higher injection S/Rs, the false-negative rate needs to be controlled to improve the precision. These factors were partially heuristically set on a separate training set of ten injections.
}
\end{table}

\subsection{Classical aperture photometry}\label{cap}
Classical aperture photometry (CAP) was performed using the Python toolbox Photutils \citep{bradley2016photutils}. The aperture and annulus settings are given in Table \ref{tab:aperture_params}.
\begin{table}[htbp]
\renewcommand{\arraystretch}{1.5}
    \caption{\label{tab:aperture_params}Aperture photometry parameters.}
    \centering
    {\small
    \begin{tabular}{|l|c|c|}
        \hline
         \rowcolor[gray]{0.9} Parameter & VLT/VISIR & SOFIA/FORCAST \\
        \hline \hline
        Aperture size & 39 pixels (1.76\arcsec) & 15 pixels (11.52\arcsec) \\
        \hline
        Annulus inner size & 39 pixels (1.76\arcsec) & 17 pixels (13.06\arcsec) \\
        \hline
        Annulus outer size & 69 pixels (3.10\arcsec) & 30 pixels (23.04\arcsec) \\
        \hline
    \end{tabular}
    }
\tablefoot{
Parameters used for the VISIR and SOFIA datasets. Measurements include the central aperture size and the inner and outer sizes of the background annulus, specified in pixels and arcsecs. Aperture and annulus are both square. These values are based on measured source sizes in the data. For VLT/VISIR, we used a value of $3\times\text{FWHM}$ to ensure that almost all flux ({$99.9~\%$ of the total flux, relatively speaking}) was captured for a reliable measurement of the photometric error.
}
\end{table}

This geometric configuration was empirically set for the VISIR and SOFIA diffraction-limited PSF and mid-infrared sky-background characteristics. The source positions were known a priori for VLT/VISIR and were manually set for SOFIA/FORCAST.

The classical aperture photometry (CAP) is defined as
\begin{equation}
\label{eq:S/R3_discrete_2}
\hat{F}_{\text{source}} = \sum_{t\in n_t} \; \sum_{(x,y) \in A} \left[ I(x,y,t) - \mu_{\text{annulus}} \right],
\end{equation}
where $I(x,y,t)$ is the flux density at position $(x,y)$ at time $t$, $A$ denotes the spatial area of the considered source, defined by the aperture, and $\mu_{\text{annulus}}$ is the median flux density in the annulus. $\hat{F}$ is the estimated source flux density. For chop-only, we applied CAP on the chop time-frame average, for the chop-nod data we applied CAP on the chop-nod time frame average, and for LORABEL, we applied CAP on the $C$ time-frame average. For all methods, we kept the integration time equal. 

\subsection{Metrics}
We have two main metrics: the photometric error, and the S/R. The photometric error ($\phi$) is defined as
\begin{equation}
\label{eq:phot_error}
\begin{aligned}
\phi = \frac{\hat{F}_{\text{source}}-F_{\text{source}}}{F_{\text{source}}}\times 100 \%,
\end{aligned}
\end{equation}
where $\hat{F}_{\text{source}}$ is the estimated source flux density from CAP after background removal, and $F_{\text{source}}$ is the injected source flux. We used a simple S/R relation, defined as
\begin{equation}
\label{eq:S/R2}
\begin{aligned}
\text{S/R} = \frac{\hat{F}_{\text{source}}}{\sqrt{n_{\text{pix}}\cdot \sigma_{\text{bkg}}^2}},
\end{aligned}
\end{equation}
where we assumed no read-out noise, no dark current noise, and no photon shot noise associated with the point source. {$n_{\text{pix}}$ is considered the number of pixels used in the aperture photometry, as defined in Table \ref{tab:aperture_params}}. Finally, for the detection precision ($\pi$), we used the following definition:
\[
\pi = \frac{\text{TP}}{\text{TP}+\text{FP}},
\]
where $\text{TP}$ (true positives) counts the correct detections, and $\text{FP}$ (false positives) counts the incorrect detections.

\section{Method}
\begin{figure}
\centering
  \includegraphics[width=\linewidth]{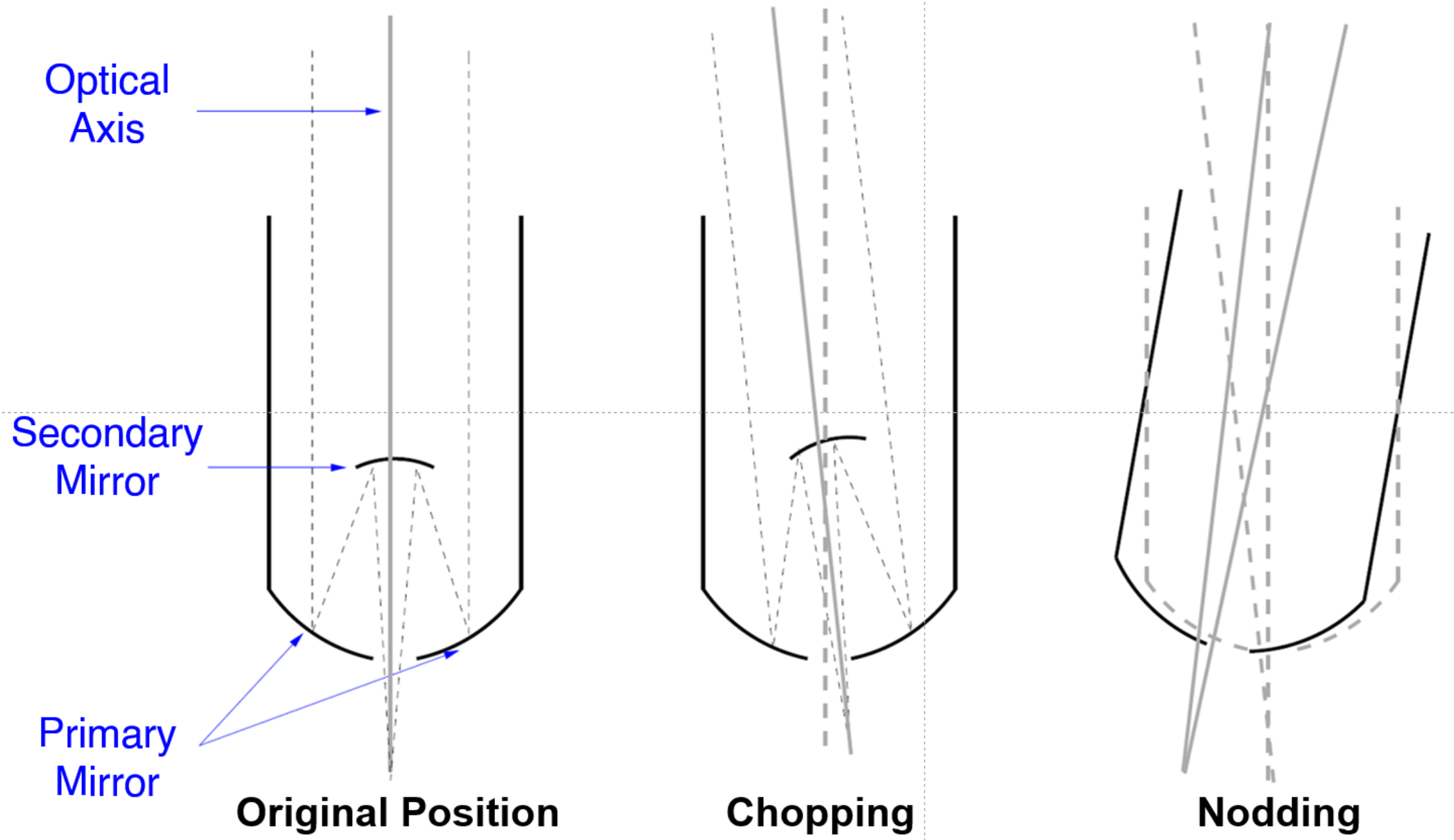}
  \caption{Abstract visualisation of chopping and nodding, adapted from \citet{lenzen2006statistical}. Left panel: Original position of the telescope and secondary mirror. Middle: Chopping technique, where the secondary mirror is rapidly tilted back and forth to observe the target and a nearby reference position. Right: Nodding technique, where the entire telescope assembly is periodically tilted to observe the target and a nearby reference position.}
  \label{fig:chop_nod}
\end{figure}

We assumed that a measurement at chop position $j$, time step $t$, and one particular nod position $\alpha$ or $\beta$, $X^{j,\alpha}_t \in \mathbb{R}^{m \times n}$, consists of the source signal $S^j_t \in \mathbb{R}^{m \times n}$ and an additive background signal $E_t^{j,\alpha} \in \mathbb{R}^{m \times n}$, such that we can model it as

\begin{equation}
\label{eq:base}
\begin{aligned}
X^{j,\alpha}_t = S^{j,\alpha}_t + E^{j,\alpha}_t.
\end{aligned}
\end{equation}
For each chop position, the source moves in the focal plane due to a tilt of the secondary mirror. With nodding, the entire telescope structure is moved, such that point sources also shift, and chop positions will move sources in different directions relative to the time frame (see Fig. \ref{fig:chop_nod}). An example image is given in Fig. \ref{fig:visir_overview_image} (VLT/VISIR), where we observe a positive and negative point source that make up $S^{j,\alpha}_t$, and the background signal $E_t^{j,\alpha}$ that appears as a gradient from bottom left to top right.

\subsection{Traditional chop-nod subtraction}
The traditional way to perform a background subtraction is by performing a simple chop-nod scheme. These techniques are used to mitigate the effects of strong and variable background radiation in mid-infrared astronomy. The chop sequence is performed with the smaller secondary mirror, which allows for fast offsets to track variability of the background on sub-second timescales. However, the tilt of the secondary mirror breaks the symmetry of the optical system (with respect to the optical axis) and introduces residuals, which require a mirrored observing sequence. The offset for this mirrored sequence is performed with the primary mirror. This nodding with the larger and more massive primary mirror takes more time, which is acceptable because the above-mentioned residuals vary on a much slower timescale of minutes. We considered four different measurements, that is, two measurements at different chop positions ($X^1_t$ and $X^2_t$) and two measurements at different nod positions (e.g. $X^{1,\alpha}_t$ and $X^{1,\beta}_t$) for each chop position. First, we obtained a chop subtraction for each nod position by simple subtraction of the two chop images for each nod position,
\begin{equation}
\label{eq:0}
\begin{aligned}
X^\alpha_t &= X^{1,\alpha}_t-X^{2,\alpha}_t \\
&= \underbrace{(S^{1,\alpha}_t - S^{2,\alpha}_t)}_{\text{chop sources:}\ S^\alpha_t} + \underbrace{(E^{1,\alpha}_t - E^{2,\alpha}_t)}_{\text{chop residual:}\ \Delta E^{\alpha}_t}.
\end{aligned}
\end{equation}

We obtained two copies of the original source, one positive and one negative pattern. To further reduce imperfections left by the chop subtraction, so-called chop residuals and denoted by $\Delta E^{\alpha}_t$ and $\Delta E^{\beta}_t$, nod subtraction was applied. The resulting chop-nod subtraction at time step $t$, that is, $Y_t$, can then be formulated as
\begin{equation}
\label{eq:1}
\begin{aligned}
Y_t &= \underbrace{\overbrace{(X^{1,\alpha}_t-X^{2,\alpha}_t)}^{\text{chop subtraction}}-\ (X^{1,\beta}_t-X^{2,\beta}_t)}_{\text{chop-nod subtraction}} \\
&= \underbrace{(S^\alpha_t-S^\beta_t)}_{\text{chop-nod sources:}\ S_t} + \underbrace{(\Delta E^{\alpha}_t - \Delta E^\beta_t)}_{\text{chop-nod residual:}\ \Delta E_t}.
\end{aligned}
\end{equation}

We then obtained {four} source patterns in our chop-nod subtraction $Y_t$ of the original source, {generally} two negatives and one positive source, where the positive source has double intensity with respect to the negative sources, {when the nodding offset is parallel and equal in amplitude to the chopping}. This scheme is usually individually applied on all available time frames and then subsequently time averaged, for example to reduce for photon shot noise, that is, $Y = \frac{1}{n_t}\sum_t Y_t$, where $n_t$ is the total number of available time frames. This method exploits source signal shifts relative to the background signal to prevent source subtraction or removal. The assumption is that (1) a large part of the background signal in each chop position is similar (i.e. remains constant because the chop rate is fast), and (2) that chop residuals (here considered residual background signal) in two different nod positions are similar in nature. No mathematical (structural) properties of the background are therefore exploited. This scheme starts from sub-noise sources and is able to result in 
detectable sources, with an S/N high enough for safe use in the subsequent analysis.

\subsection{Proposed method: LOw-RAnk Background ELimination (LORABEL)}\label{sec:proposed}
For LORABEL, we require a time series of $n_t$ chopped/chop-only frames (i.e. no nodding). Ideally, the frames are sufficiently integrated\footnote{as a rule of thumb, the structural residual patterns should be visible in individual frames, not purely randomly distributed noise/photon shot noise}. This is largely dependent on the instrumental set-up, experimental design, and pre-processing pipeline. The number of chopped frames $n_t$ required for the working of LORABEL depends on the complexity (rank) of the background structure\footnote{as a rule of thumb, we take at least a tenfold number of frames with respect to the rank}. The latter can be estimated by performing an SVD on an empty dataset and examining the leading singular values. In general, more frames are better for a more accurate noise reduction. \par
\begin{figure*}[h!]
\centering
  \includegraphics[width=\linewidth]{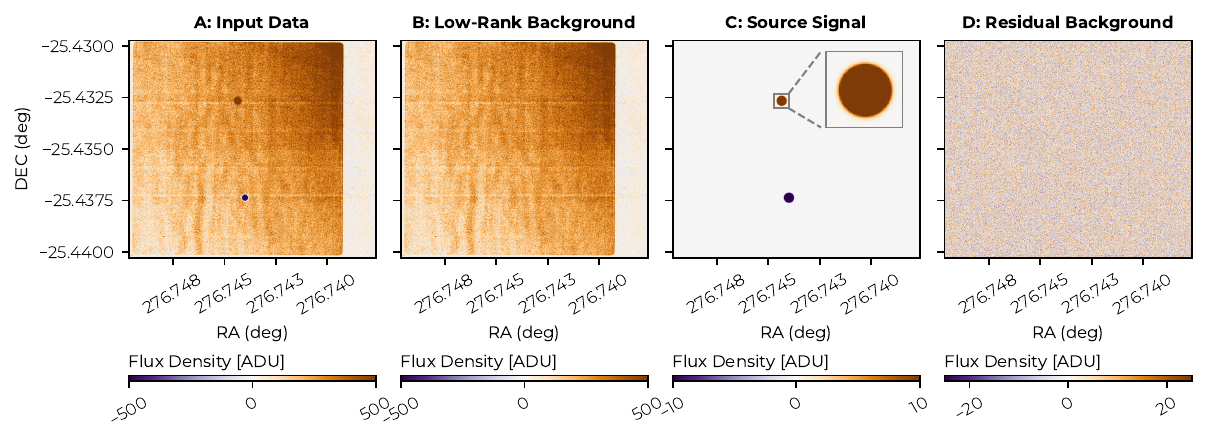}
  \caption{Ideal decomposition of the input image with very high injection S/R (A) into three components: (B) low-rank thermal background, (C) injected source signal with positive and negative sources, and (D) Gaussian noise. The combined model follows A = B + C + D, illustrating the contribution of each component to the final observed data. These images are on a different intensity scales.}
  \label{fig:plot6}
\end{figure*}

 The method we propose is based on a low-rank background signal and sparse source signal model. It therefore acts on the chop-only time frames of point sources, that is, $X_t = X^1_t-X^2_t$ (for simplicity, we omit the $\alpha$ and $\beta$ nod position notation as we have only a single position $\alpha$ or $\beta$), where we assumed that the chop residuals (i.e. residual background signal, $\Delta E_t$) remains (nearly) constant over different time frames. To apply LORABEL, we first grouped the chop subtractions of the available time series as follows:
\begin{gather}
\begin{split}
\label{eq:grouping}
X = \begin{bmatrix} \text{vec}(X_1) & \text{vec}(X_2) & \cdots & \text{vec}(X_{n_t})
\end{bmatrix} \in \mathbb{R}^{mn \times n_t}.
\end{split}
\end{gather}
Each column in the matrix $X$ consists of a vectorised chop subtraction at time step $j$, while each row represents a time series of a single imaging detector pixel (that was also chop subtracted). Using Eq. \ref{eq:0}, we decomposed our matrix as
\begin{gather}
\begin{split}
\label{eq:representation}
X = S + \Delta E,
\end{split}
\end{gather}
where $S \in \mathbb{R}^{mn \times n_t}$ contains the point-source signal, and $\Delta E \in \mathbb{R}^{mn \times n_t}$ consists of quasi-static instrumental patterns (i.e. low rank), slowly variable background emission patterns of low spatial frequency (i.e. low rank), and highly time-variable photon shot noise (considered to be Poisson/Gaussian noise). The model we propose can therefore be considered a member of the family ${\cal{F}}$ of extended linear models,
\begin{gather}
\begin{split}
\label{eq:2}
{\cal{F}}: \ A = B + C + D,
\end{split}
\end{gather}
where $A \in {\mathbb{R}}^{mn \times n_t}$ represents a measurement matrix that is decomposed into $B \in {\mathbb{R}}^{mn \times n_t}$, $C \in {\mathbb{R}}^{mn \times n_t}$, and $D \in {\mathbb{R}}^{mn \times n_t}$, which are matrix terms differing in mathematical properties. We defined them as a low-rank term ($B$), a sparse term ($C$), and a residual term ($D=A-B-C$). For our purposes, $B$ and $D$ are expected to model the content of $\Delta E$, as we respectively assumed the quasi-static instrumental patterns and slowly variable background emission patterns of low spatial frequency to be low-rank and highly time-variable photon shot noise to be modelled as small dense noise (see mathematical notations). At the same time, the point sources were considered to be spatially sparse even though a second copy was created during chopping, and we can therefore model them by $C$. Figure \ref{fig:plot6} shows this decomposition of the input data (A) into three components: (B) a low-rank thermal background, (C) a source signal containing both positive and negative sources, and (D) Gaussian i.i.d. noise. This decomposition illustrates the interaction of the various signal components (with different flux densities), with (B) the low-rank background representing a large-scale thermal structure (considered noise), (C) a source signal capturing the astrophysical objects of interest, and (D) a randomly distributed noise component simulating observational uncertainties.

\subsubsection{Point sources: Low rank and sparse}\label{lowrank}
In an idealised setting, aside from their inherent spatial sparsity, point source signals would appear as a rank-$1$ feature in the time series data, as presented in Eq. \ref{eq:grouping}. However, since the flux of these point sources is typically much lower than the overall background flux, especially when their S/R is low, their corresponding component is relatively weak in an $\ell_2$-sense and does not dominate the leading components that constitute the $B$ term. A key assumption here is that the temporal components of the point sources remain uncorrelated with the quasi-static instrumental and slowly varying, low-spatial-frequency background emissions represented by the $B$ term. Our first case study reveals that this assumption can be violated, causing some point-source signal to leak into the $B$ term. A similar effect was observed in an exoplanet detection for a similar class of methods \citep{pueyo2016detection}. In real-world observations, such as those from SOFIA/FORCAST, factors such as atmospheric jitter, variations in instrument response, and imperfect pre-processing can alter this rank-$1$ property of point sources because its spatial shape, location, and temporal pattern can alter in a non-linear fashion. Instead of remaining confined to a rank-$1$ representation, point sources tend then to {spread} across multiple components and are thought to become less strongly correlated to the stable low-rank background. Interestingly, this effect might even lead to an improved retrieval of the source flux density compared to what the idealised model predicts. This property is exploited by some drift-scanning techniques \citep{ohsawa2018slow}. LORABEL therefore does not require the background to be perfectly stable throughout the entire dataset. The method assumes that over a short enough time interval, the background is locally stable, which allows us to distinguish features from shot noise, for instance. Small and gradual background changes are acceptable. By ``slowly changing,'' we mean variations that occur on timescales longer than the characteristic duration needed to establish the static baseline of features, typically between 10 and 100 frames, depending on the complexity of the background. At the same time, we here focused on chopped data because this observing mode is most relevant for METIS and is technically best suited for testing LORABEL at this stage. In the N band, ground-based observations are strongly dominated by background, and without chopping, the data would accumulate significant detector noise. LORABEL is not designed to specifically mitigate detector noise. Its strength lies in handling structured backgrounds and source extraction. For this reason, applying LORABEL to non-chopped images would introduce challenges unrelated to the algorithm itself, and would not provide a fair evaluation of its performance.

\subsubsection{Optimisation}
The above problem of decomposing low-rank, sparse, and small dense terms can be solved by stable principal component pursuit \citep[SPCP, ][]{zhou2010stable}. Zhou et al. (2010) provided conditions related to (1) the incoherence of bases \citep{candes2007sparsity}, (2) the rank of the low-rank term $B$, (3) the sparsity of the sparse term, $C$, and (4) the magnitude of the dense term, $||A-B-C||_F$, which leads to an approximate solution (close to the exact solution). We assumed that these conditions are valid for our problem. The optimisation program defined in Eq. \ref{eq:SPCP} is proposed to solve this problem { (see Sect. \ref{maths_notations} for mathematical notations)}. We denote stable principal component pursuit (SPCP) as $f_{\text{SPCP}}:{\mathbb{R}}^{m \times n}\times{\mathbb{R}}^{+}\times{\mathbb{R}}^{+}\rightarrow{\mathbb{R}}^{m \times n}\times{\mathbb{R}}^{m \times n}; (A,\theta,\delta) \mapsto (B,C)$,

\begin{equation}
\label{eq:SPCP}
\begin{aligned}
&\underset{B, C}{\text{minimise}} && \nuclearnorm{B}+\theta\onenorm{C}\\
&\text{subject to} && \frobeniusnorm{A-B-C}\leq \delta.\\
\end{aligned}
\end{equation}
We observe that an input matrix $A$ should be provided along with two parameters, $\theta$ and $\delta$. $\theta$ trades off the sparsity of the matrix $C$ with the rank of $B$ (i.e. a trade-off between source signal retention and background signal reduction), while $\delta$ acts as a cut-off related to noise. For i.i.d. element-wise, Gaussian noise with mean zero and standard deviation $\sigma$, $\delta$ is defined as $\delta=\sqrt{mn}\sigma$.\par
Moreover, we observe that for background modelling, we do not have to explicitly deal with bad pixels, for instance caused by pixel saturation, cosmic rays, or satellite flybys because they would naturally be filtered and end up in the $C$ term as long as they occur as sparse and not coherent phenomena. Additionally, this model does not require masking, removing point sources, or having information about dithering settings. We note that incorporating a priori information into dithering or beam positions could improve performance, which we leave as a potential direction for future work.

\subsubsection{Parameter tuning}\label{sec:param}
As previously outlined in Eq. \ref{eq:SPCP}, the model involves two key parameters that require tuning: $\delta\in\mathbb{R}$ and $\theta\in\mathbb{R}$. The parameter $\delta$ can be estimated by the standard deviation of the i.i.d. Gaussian noise assumed to be present in the data. Meanwhile, an initial estimate for $\theta$ is given by $\theta=\sqrt{1/\max{(m,n)}}$, where $m,n$ are the matrix sizes of the input matrix $A\in\mathbb{R}^{m \times n}$ \citep{zhou2010stable}. These estimates assume that ideal conditions are met. At the same time, it has been suggested that fine-tuning these parameters can yield improved results \citep{candes2011robust}. \par 
For the first case study, we therefore performed a grid search between $0.1\times \bar{\theta} \leq \theta \leq 1.1\times\bar{\theta}$ where $\bar{\theta}=\sqrt{1/\max{(m,n)}}$ and fixed $\delta=\frobeniusnorm{E}$, where $E$ is the matrix assumed exclusively containing i.i.d. Gaussian noise (known a priori), as by definition $\frobeniusnorm{E}=\sigma^2 mn$, where $\sigma$ is defined as the standard deviation of the noise matrix $D$. For detection, these parameters were slightly adjusted to a $\theta=1.1\times\sqrt{1/\max{(m,n)}}$ and a $\delta=.7\times\frobeniusnorm{A}$. This ensured that most of the low-rank background was removed and that source signal was retained as much as possible. {The required runtime per parameter setting was around $30$~s.}\par 
For the second case study, we conducted a parameter-tuning exercise by performing a grid search over a $100\times 100$ parameter setting range for both parameters around $\theta=\sqrt{1/\max{(m,n)}}$ and $\delta=\frobeniusnorm{A}$ (we note that $A$ is the input matrix, as $E$ is not known a priori). This grid search was executed using two A6000 NVIDIA GPUs to accelerate the process. The optimal parameter set was selected based on the criteria of maximum background reduction and maximum retention of the source signal by performing aperture photometry for each parameter setting. {We therefore equally weighted the average background-subtracted source flux and the average background flux, providing a balanced measure of background suppression and photometric fidelity. The equal weighting was chosen to avoid bias towards either metric and to reduce the risk of overfitting.}Heuristically, we found that setting the search space to $0.1\times\sqrt{1/\max{(m,n)}}\leq\theta\leq\sqrt{1/\max{(m,n)}}$ and $10^{-10}\times mn\leq\delta\leq 10\times mn$ was most efficient. {We furthermore chose to linearly explore the $\theta$ space and the $\delta$ space in a logarithmic fashion. We found that values around $\bar{\theta}\approx \sqrt{1/\max{(m,n)}}$ and $\delta \approx10^{-5} m n$ were optimal, with slight variations per filter (hence the approximation).} {The required runtime per parameter setting was around $4$~s.}
As a rule of thumb, when increasing $\theta$, more background flux is removed and less source signal is kept, while lowering $\theta$ removes less background flux, but retains more source signal. The parameter $\delta$ affects both. When increasing $\delta$, it allows for more background subtraction, but probably leads to a lower source flux, while lowering $\delta$ most of the time results in more source flux, but also in more background flux.

\subsection{Availability of algorithms and data}
All methods were implemented in object-oriented Python and bundled in our LORABEL package \footnote{\url{https://github.com/vandeplaslab/lorabel}}. The data we used are public and can be found in the ESO and SOFIA archives or can be requested by contacting the authors.

\section{Results and discussion}
In the first case study, our LORABEL method was compared to the chop-only and chop-nod method on ground-based data with injected sources. This enabled an absolute evaluation for accurate source flux measurement and source detection. In the second case study, we compared the proposed method to a chop-only and chop-nod method on airborne SOFIA data. This enabled an evaluation in a practical setting at different wavelengths and with a different telescope structure. The overall goal of both case studies was to research the ability of the method in ideal circumstances, that is, with injected sources, and in practical settings.

\subsection{Quantitative study of ground-based VISIR data with injected sources of varying intensity}
\begin{figure}[!ht]
\centering
  \includegraphics[width=\linewidth,]{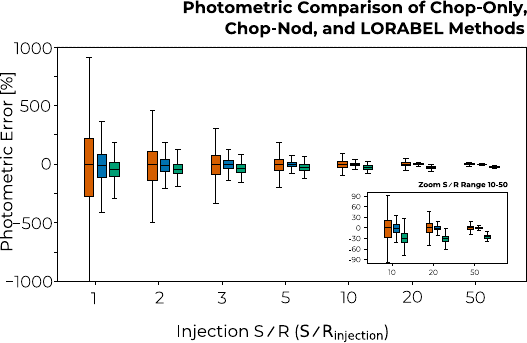}
  \caption{Comparison of the photometric error using the LORABEL method we propose, the chop-only method, and the chop-nod method at different injection S/N levels on VISIR data as specified in Table \ref{table:instrument_specs}. For each signal-to-noise ratio (S/R), the coloured rectangular boxes represent the interquartile range (25th–75th percentile) of the photometric error distribution, while the horizontal black line inside each box indicates the median value. The vertical lines (whiskers) extend to the most extreme values within 1.5x the interquartile range, following the standard boxplot definition; outliers are not shown. The y-axis displays the photometric error in percentage.}
  \label{fig:plot1}
\end{figure}

The photometric error for the three methods with different injected source flux densities (expressed as a function of the injection S/R, $\text{S/R}_\text{injection}$) is given in Fig. \ref{fig:plot1}. It consists of the percentage difference between the injected source flux and recovered source flux of the positive point source with respect to the injected source flux. We used classical aperture photometry (see Sect. \ref{cap}) to obtain the recovered source flux. The experiment was performed on $300$ different injection locations, where the local noise statistics change and thus the injection, to assess robustness. Finally, we kept the integration time for all methods equal. The following observations were made:
\begin{enumerate}
    \item Injection S/R $\leq 5$: at very low injection S/Rs, LORABEL shows a considerably smaller distribution of photometric errors than for the chop-only and chop-nod methods. Its boxplot indicates a smaller interquartile range and less pronounced outliers, suggesting that LORABEL is less susceptible to noise fluctuations in this regime. In contrast, the chop-only and chop-nod methods tend to exhibit a wider distribution, implying a less consistent performance when the source signal is very weak.
    \item Injection S/R $> 5$: for higher injection S/R values, the LORABEL performance improves slowly. Its error distribution becomes much tighter, but is caught up by chop-nodding.
    \item LORABEL bias: the median photometric error for LORABEL does not centre around zero, but is instead systematically shifted to $\approx-30\%$. This negative median indicates a systematic underestimation of the recovered source flux density relative to the injected flux.
    \item Decrease in error with increasing injection S/R: all methods exhibit a general decrease in photometric error as the injection S/R increases. This trend is expected because at high source flux densities, the background contribution becomes relatively negligible.
\end{enumerate}
The {underestimation of the photometric flux} observed in Fig. \ref{fig:plot1} is thought to be caused by the injection of a perfect low-rank source, which in reality is probably partially reduced (see Sect. \ref{lowrank}). This bias further depends on the data, that is, on instrument type and observation specifics, as well as on the parameter settings of LORABEL. As the bias appears to be systematic and only weakly related to the injection S/R, we might estimate this bias by synthetically by injecting one or multiple sources with known source fluxes, estimating this bias, and compensating for it. This would only be a solution in the case of perfect low-rank sources, such as our injections. In the case of {non-idealised point sources, that is, sources deviating from the Gaussian reference}, we might bound spatial point sources shapes and temporal patterns and search for lower and upper bounds on the resulting variation in bias. This is, however, considered to be beyond the scope for this paper.\par
\begin{figure}[h!]
\centering
\includegraphics[width=\linewidth]{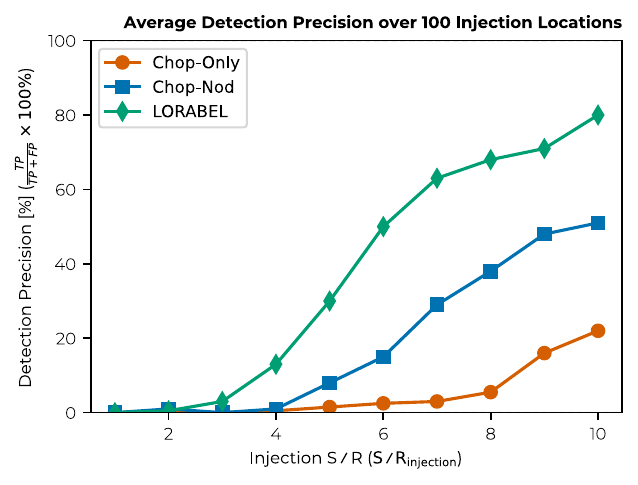}
\caption{Comparison of the detection precision among the traditional chop-only and chop-nod methods and the LORABEL approach we propose {for VISIR data}. The figure presents the average detection precision computed over $100$ random injection locations, where the local noise statistics varied with each injection. To avoid bias from selective parameter tuning, we employed a fixed set of detection parameters (see Sect. \ref{detection}) derived from a training set. The constraints roundness1, roundness2, and peak S/R were kept identical for all methods to reduce false positives, which resulted in sub-optimal outcomes as they could be tuned for each individual method. Furthermore, the use of local noise statistics for the injection and global noise statistics for detection threshold (see Sect. \ref{detection}) further leads to a reduced detection precision. Finally, we note that this detection precision reaches $100\%$ for high S/Rs, not considered relevant for this experiment.}
\label{fig:precision}
\end{figure}

 
The notably reduced {dispersion} in photometric error for LORABEL at low S/R values potentially greatly affects the detection rate of very faint sources that other methods may not find. To evaluate the detection performance, we conducted a comparative detection experiment in which data processed by all three methods were subsequently analysed using our detection algorithm (see Sect. \ref{detection}). The superior performance of the LORABEL method is demonstrated clearly in Fig. \ref{fig:precision}, where LORABEL achieves consistently higher precision at all tested injection S/R levels. The displayed precisions represent averages calculated over $100$ randomly selected injection locations, using a uniform set of algorithm parameters for the locations. It is important to note that using a fixed parameter set {for the detection algorithm}, rather than optimising parameters individually for each injection site, occasionally leads to scenarios in which no detection is achieved (resulting in true positives, TP $= 0$) or an increase in false positives (FP), thereby reducing the overall precision. In conclusion and as an important statement to the LORABEL user: the observed photometric bias is only weakly dependent on S/R and is primarily determined by the method and its parameters; LORABEL at this point in time is intended for detection and not for an accurate flux measurement.

\subsection{Comparative study with the traditional chop-nod method on airborne SOFIA data}
We evaluated the versatility of LORABEL by comparing it with traditional chop-nod techniques using airborne SOFIA observations. For this dataset, optimising parameters (see \ref{sec:param}) proved challenging because there were not enough time frames {(see Table \ref{table:instrument_specs})}. In addition, the parameter setting was strongly affected by the dataset characteristics. Therefore, we combined the approach described above with heuristic tuning to achieve reasonable parameter settings.
In Table \ref{tab:spectral_lens_comparison} the photometric analysis across multiple spectral filters reveals several key performance characteristics for LORABEL. The aperture photometry yields comparable mean flux values between LORABEL and the traditional chop-nod technique for most spectral filters, except in F111, where the chop-nod approach exhibits significant variability ($\sigma = 3211.2$ counts), likely due to temporal source misalignment. In contrast, background estimation with LORABEL proves to achieve superior stability across all bands ($\mu = -0.002$, $\sigma = 0.139$), substantially outperforming the chop-only and chop-nod approaches. The S/N analysis further demonstrates an enhanced performance relative to traditional chop-nodding across all filters, with particular improvements in F088 and F111. Although the S/R in F197 is slightly reduced because the aperture flux is lower, it remains competitive. These results validate the practical applicability of LORABEL to airborne infrared observations, successfully extending its utility from theoretical framework to real observational data.
\begin{table*}[h!]
\renewcommand{\arraystretch}{1.5}
\centering
\caption{\label{tab:spectral_lens_comparison}
Comparison of different spectral filters for the chop-only, traditional chop-nod, and LORABEL method.
}
\setlength{\tabcolsep}{6pt}
\begin{tabular}{|l|c|c|c|c|c|}
  \hline
  \rowcolor[gray]{0.9}
  Method & Filter 
    & Aperture $\mu\pm\sigma\%$ 
    & Peak $\mu\pm\sigma\%$ 
    & Annulus $\mu\pm\sigma\%$ 
    & S/R \\
  \hline
  \multirow{3}{*}{Chop-Only}  
    & F088 & $4736.8 \pm 0.83\%$  & $394.7 \pm 7.29\%$  & $0.133 \pm 630\%$   & $376.8$ \\
    & F197 & $1338.4 \pm 4.45\%$  & $90.5 \pm 7.07\%$   & $-0.095 \pm 486\%$  & $193.1$ \\
    & F111 & $3353.8 \pm 19.8\%$  & $254.1 \pm 11.8\%$  & $-0.516 \pm 206\%$  & $209.7$ \\
  \hline
  \multirow{3}{*}{Chop-Nod}  
    & F088 & $4227.6 \pm 1.45\%$  & $386.4 \pm 4.32\%$  & $0.100 \pm 1051\%$  & $268.2$ \\
    & F197 & $877.1 \pm 7.10\%$   & $93.0 \pm 4.62\%$   & $0.500 \pm 125\%$   & $93.4$  \\
    & F111 & $2425.5 \pm 132\%$   & $308.3 \pm 17.2\%$  & $-1.000 \pm 198\%$  & $81.9$  \\
  \hline
  \multirow{3}{*}{LORABEL}  
    & F088 & $4247.0 \pm 2.82\%$  & $391.3 \pm 7.28\%$  & $-0.002 \pm 6950\%$ & $445.9$ \\
    & F197 & $1040.0 \pm 8.13\%$  & $87.1 \pm 7.69\%$   & $-0.001 \pm 2600\%$ & $176.0$ \\
    & F111 & $2931.0 \pm 18.6\%$  & $248.6 \pm 11.2\%$  & $0.032 \pm 809\%$   & $880.2$ \\
  \hline
\end{tabular}
\tablefoot{
Comparison of aperture, peak, annulus, and S/R statistics on SOFIA data as specified in Table \ref{table:instrument_specs}. The reported averages ($\mu$) and standard deviations ($\sigma$) were obtained with classical aperture photometry by considering the time-frame average of each of the four available repeated measurements as a separate realisation (see Table~\ref{table:instrument_specs}). 'Aperture' refers to the source flux density, 'Peak' denotes the maximum pixel brightness detected by the DAOStarFinder routine in the aperture, and 'Annulus' reflects the mean flux density in the annulus. The S/R in this table was obtained by averaging the standard deviation of the source flux density and background flux density  over all four repeated measurements. The errors are reported in relative terms for a consistent comparison at different count levels.
}
\end{table*}

Visual inspection of the results, as shown in the decomposition of the time-integrated SOFIA F197 spectral filter in Fig. \ref{fig:plot5}, demonstrates that LORABEL effectively separates the background, source signal, and noise components. Although the source signal component ($C$) does not exhibit a perfectly flat background across all pixels, it nonetheless facilitates a clearer analysis of the source under challenging observational conditions. This is further confirmed by the examination of the background distribution around the source, as illustrated in Fig. \ref{fig:plot7}.

\section{Conclusions}
We introduced a novel computational technique, LORABEL, for background subtraction in mid-infrared astronomy, and benchmarked its performance against chop-only and traditional chop-nod methods. Based on simulated VISIR data with injected point sources as well as real SOFIA observations, our results revealed several key findings.
For the VISIR case study, LORABEL delivered more consistent flux measurements in regimes with a low signal-to-noise ratio compared to the conventional methods. Although the method systematically underestimates the source flux, which suggests that a portion of the true signal is inadvertently removed, it substantially reduces variability in the photometric error. While the percentage of flux loss is dependent on the parameter settings and is specific to a dataset, a way to compensate for this loss in a real-world setting might be the introduction of artificial sources with a known source flux in a dataset. Furthermore, we showed that LORABEL is particularly promising for the detection of faint sources under challenging conditions. We showed in our second experiment that LORABEL outperforms the chop-only and chop-nod methods in detection precision for all S/R cases. 
In the airborne SOFIA dataset, LORABEL achieved a reduction in the mean background flux (by factors of $3$ to $10$) while largely preserving the source signal, leading to improved signal-to-noise ratios at all spectral bands with respect to the traditional chop-nodding method. Despite a slight increase in the standard deviation of the source flux measurements, the results confirm that LORABEL can perform even in non-ideal scenarios with limited time frames.
While the optimisation of LORABEL can be challenging due to its complex parameter space and sensitivity to various observational factors, our results clearly show that effective background subtraction is achievable without relying on source masking or additional nodding.
In summary, LORABEL shows significant potential for enhancing mid-infrared observations, especially for source detection in low-S/N regimes when nodding is not available. Nonetheless, further refinements are required to mitigate its systematic flux underestimation and to fully optimise its performance under varying observational conditions.

\begin{acknowledgements}
We thank Olivier Absil for his valuable feedback improving our manuscript.

RARM is supported by a TU Delft Space Institute Seed Grant. 

AP is supported by the {Deut\-sche For\-schungs\-ge\-mein\-schaft, DFG\/} project number PI 2102/1-1

Based on observations collected at the European Southern Observatory under ESO programme 60.A-9800(l).

Based in part on observations made with the NASA/DLR Stratospheric Observatory for Infrared Astronomy (SOFIA). SOFIA is jointly operated by the Universities Space Research Association, Inc. (USRA), under NASA contract NNA17BF53C, and the Deutsches SOFIA Institute (DSI) under DLR contract 50 OK 2002 to the University of Stuttgart.

We thank Bill Vacca for his helpful coordination and support in arranging the technical FORCAST observations at the SOFIA telescope.

This research has made use of NASA's Astrophysics Data System (ADS) bibliographic services. 

We acknowledge the community efforts devoted to the development of the following open-source packages that were used in this work: numpy (\href{http:\\numpy.org}{numpy.org}), matplotlib (\href{http:\\matplotlib.org}{matplotlib.org}), and astropy (\href{http:\\astropy.org}{astropy.org}). Research reported in this publication was supported by the National Institutes of Health (NIH)’s Common Fund, National Institute Of Diabetes And Digestive And Kidney Diseases (NIDDK), and the Office Of The Director (OD) under awards U54DK120058, U54DK134302, and U01DK133766 (RV), by NIH’s Common Fund, National Eye Institute, and the Office Of The Director (OD) under award U54EY032442 (RV), by NIH’s National Institute Of Allergy And Infectious Diseases (NIAID) under awards R01AI138581, 2R01AI138581, and R01AI145992 (RV),  and by NIH’s National Institute On Aging (NIA) under award R01AG078803 (RV). The content is solely the responsibility of the authors and does not necessarily represent the official views of the National Institutes of Health.
\end{acknowledgements}

\bibliographystyle{aa}
\bibliography{Bibliography}

@article{lacy2002texes,
  title={TEXES: A sensitive high-resolution grating spectrograph for the mid-infrared},
  author={Lacy, JH and Richter, MJ and Greathouse, TK and Jaffe, DT and Zhu, Q},
  journal={\pasp},
  volume={114},
  number={792},
  pages={153},
  year={2002},
  publisher={IOP Publishing}
}

@misc{esoOverview,
	author = {ESO},
	title = {{E}{S}{O} - {O}verview --- eso.org},
	howpublished = {\url{https://www.eso.org/sci/facilities/paranal/instruments/visir/overview.html}},
	year = {2024},
	note = {[Accessed 27-02-2025]},
}

@INPROCEEDINGS{Telesco2003,
       author = {{Telesco}, Charles M. and {Ciardi}, David and {French}, James and {Ftaclas}, Christ and {Hanna}, Kevin T. and {Hon}, David B. and {Hough}, James H. and {Julian}, Jeffrey and {Julian}, Roger and {Kidger}, Mark and {Packham}, Chris C. and {Pina}, Robert K. and {Varosi}, Frank and {Sellar}, R. Glenn},
        title = "{CanariCam: a multimode mid-infrared camera for the Gran Telescopio CANARIAS}",
    booktitle = {Instrument Design and Performance for Optical/Infrared Ground-based Telescopes},
         year = 2003,
       editor = {{Iye}, Masanori and {Moorwood}, Alan F.~M.},
       series = {Society of Photo-Optical Instrumentation Engineers (SPIE) Conference Series},
       volume = {4841},
        month = mar,
        pages = {913-922},
          doi = {10.1117/12.458979},
       adsurl = {https://ui.adsabs.harvard.edu/abs/2003SPIE.4841..913T}
}

@article{werner2004spitzer,
  title={The Spitzer space telescope mission},
  author={Werner, Michael W and Roellig, Thomas L and Low, FJ and Rieke, George H and Rieke, M and Hoffmann, WF and Young, E and Houck, JR and Brandl, B and Fazio, GG and others},
  journal={\apjs},
  volume={154},
  number={1},
  pages={1},
  year={2004},
  publisher={IOP Publishing}
}

@article{irace1983iras,
  title={The IRAS telescope.},
  author={Irace, W and Rosing, D},
  journal={Journal of the British Interplanetary Society},
  volume={36},
  pages={27--33},
  year={1983}
}

@inproceedings{rio1998visir,
  title={VISIR: the mid-infrared imager and spectrometer for the VLT},
  author={Rio, Yvon and Lagage, Pierre-Olivier and Dubreuil, Didier and Durand, Gilles A and Lyraud, Charles and Pel, Jan-Willem and de Haas, Johannes CM and Schoenmaker, Anton and Tolsma, Hoite},
  booktitle={Infrared Astronomical Instrumentation},
  volume={3354},
  pages={615--626},
  year={1998},
  organization={SPIE}
}

@inproceedings{krabbe2000sofia,
  title={SOFIA telescope},
  author={Krabbe, Alfred},
  booktitle={Airborne Telescope Systems},
  volume={4014},
  pages={276--281},
  year={2000},
  organization={SPIE}
}

@article{rieke2015mid,
  title={The mid-infrared instrument for the james webb space telescope, i: Introduction},
  author={Rieke, George H and Wright, GS and B{\"o}ker, T and Bouwman, J and Colina, L and Glasse, Alistair and Gordon, KD and Greene, TP and G{\"u}del, Manuel and Henning, Th and others},
  journal={\pasp},
  volume={127},
  number={953},
  pages={584},
  year={2015},
  publisher={IOP Publishing}
}

@inproceedings{brandl2012metis,
  title={METIS: the thermal infrared instrument for the E-ELT},
  author={Brandl, Bernhard R and Lenzen, Rainer and Pantin, Eric and Glasse, Alistair and Blommaert, Joris and Meyer, Michael and Guedel, Manuel and Venema, Lars and Molster, Frank and Stuik, Remko and others},
  booktitle={Ground-based and Airborne Instrumentation for Astronomy IV},
  volume={8446},
  pages={554--566},
  year={2012},
  organization={SPIE}
}

@inproceedings{burtscher2020towards,
  title={Towards a physical understanding of the thermal background in large ground-based telescopes},
  author={Burtscher, Leonard and Politopoulos, Ioannis and Fern{\'a}ndez-Acosta, Sergio and Agocs, Tibor and van den Ancker, Mario and van Boekel, Roy and Brandl, Bernhard and K{\"a}ufl, Hans-Ulrich and Pantin, Eric and Pietrow, Alex GM and others},
  booktitle={Ground-based and Airborne Instrumentation for Astronomy VIII},
  volume={11447},
  pages={1678--1694},
  year={2020},
  organization={SPIE}
}

@ARTICLE{Herter2018forcast,
       author = {{Herter}, T.~L. and {Adams}, J.~D. and {Gull}, G.~E. and {Schoenwald}, J. and {Keller}, L.~D. and {Pirger}, B.~E. and {Henderson}, C.~P. and {Stacey}, G.~J. and {Nikola}, T. and {De Buizer}, J.~M. and {Vacca}, W.~D. and {Ennico}, K.},
        title = "{FORCAST: A Mid-Infrared Camera for SOFIA}",
      journal = {Journal of Astronomical Instrumentation},
         year = 2018,
        month = jan,
       volume = {7},
       number = {4},
          eid = {1840005-451},
        pages = {1840005-451},
          doi = {10.1142/S2251171718400056},
       adsurl = {https://ui.adsabs.harvard.edu/abs/2018JAI.....740005H}
}

@ARTICLE{Petit2020,
       author = {{Petit dit de la Roche}, D.~J.~M. and {van den Ancker}, M.~E. and {Kissler-Patig}, M. and {Ivanov}, V.~D. and {Fedele}, D.},
        title = "{New constraints on the HR 8799 planetary system from mid-infrared direct imaging}",
      journal = {\mnras},
         year = 2020,
        month = jan,
       volume = {491},
       number = {2},
        pages = {1795-1799},
          doi = {10.1093/mnras/stz3117},
archivePrefix = {arXiv},
       eprint = {1911.04814},
 primaryClass = {astro-ph.EP},
       adsurl = {https://ui.adsabs.harvard.edu/abs/2020MNRAS.491.1795P}
}

@INPROCEEDINGS{2014Heikamp,
       author = {{Heikamp}, Stephanie and {Brandl}, Bernhard R. and {Keller}, Christoph U. and {Venema}, Lars and {Pantin}, Eric and {Siebenmorgen}, Ralf and {Ives}, Derek and {Kerber}, Florian},
        title = "{Drift scanning technique for mid-infrared background subtraction}",
    booktitle = {Ground-based and Airborne Instrumentation for Astronomy V},
         year = 2014,
       editor = {{Ramsay}, Suzanne K. and {McLean}, Ian S. and {Takami}, Hideki},
       series = {Society of Photo-Optical Instrumentation Engineers (SPIE) Conference Series},
       volume = {9147},
        month = aug,
          eid = {91479T},
        pages = {91479T},
          doi = {10.1117/12.2065675},
       adsurl = {https://ui.adsabs.harvard.edu/abs/2014SPIE.9147E..9TH}
}

@ARTICLE{Papoular1983,
       author = {{Papoular}, R.},
        title = "{The processing of infrared sky noise by chopping, nodding and filtering}",
      journal = {\aap},
         year = 1983,
        month = jan,
       volume = {117},
       number = {1},
        pages = {46-52},
       adsurl = {https://ui.adsabs.harvard.edu/abs/1983A&A...117...46P}
}

@MASTERSTHESIS{Pietrow2016,
       author = {{Pietrow}, Alexander G.~M.},
        title = "{Mid-IR background calibrations for the E-ELT's METIS instrument}",
       school = {Leiden University},
         year = 2016,
        month = jul,
       adsurl = {https://ui.adsabs.harvard.edu/abs/2016MsT.........48P}
}

@article{pietrow2019inverse,
       author = {{Pietrow}, A.~G.~M. and {Burtscher}, L. and {Brandl}, B.},
        title = "{Inverse Chop Addition: Thermal IR Background Subtraction without Nodding}",
      journal = {Research Notes of the American Astronomical Society},
         year = 2019,
        month = feb,
       volume = {3},
       number = {2},
          eid = {42},
        pages = {42},
          doi = {10.3847/2515-5172/ab09fa},
archivePrefix = {arXiv},
       eprint = {1903.05680},
 primaryClass = {astro-ph.IM},
       adsurl = {https://ui.adsabs.harvard.edu/abs/2019RNAAS...3...42P}
}

@article{rousseau2024improving,
  title={Improving mid-infrared thermal background subtraction with Principal Component Analysis},
  author={Rousseau, H{\'e}l{\`e}ne and Ertel, Steve and Defr{\`e}re, Denis and Faramaz, Virginie and Wagner, Kevin},
  journal={\aap},
  volume={687},
  pages={A147},
  year={2024},
  publisher={EDP Sciences}
}

@inproceedings{zhou2010stable,
  title={Stable principal component pursuit},
  author={Zhou, Zihan and Li, Xiaodong and Wright, John and Candes, Emmanuel and Ma, Yi},
  booktitle={2010 IEEE international symposium on information theory},
  pages={1518--1522},
  year={2010},
  organization={IEEE}
}

@article{candes2007sparsity,
  title={Sparsity and incoherence in compressive sampling},
  author={Candes, Emmanuel and Romberg, Justin},
  journal={Inverse problems},
  volume={23},
  number={3},
  pages={969},
  year={2007},
  publisher={IOP Publishing}
}

@article{candes2011robust,
  title={Robust principal component analysis?},
  author={Cand{\`e}s, Emmanuel J and Li, Xiaodong and Ma, Yi and Wright, John},
  journal={Journal of the ACM (JACM)},
  volume={58},
  number={3},
  pages={1--37},
  year={2011},
  publisher={ACM New York, NY, USA}
}

@article{matthews2024temperate,
  title={A temperate super-Jupiter imaged with JWST in the mid-infrared},
  author={Matthews, EC and Carter, AL and Pathak, P and Morley, CV and Phillips, MW and PM, S Krishanth and Feng, F and Bonse, MJ and Boogaard, LA and Burt, JA and others},
  journal={Nature},
  pages={1--4},
  year={2024},
  publisher={Nature Publishing Group}
}

@article{wagner2021imaging,
  title={Imaging low-mass planets within the habitable zone of $\alpha$ Centauri},
  author={Wagner, Kevin and Boehle, Anna and Pathak, Prashant and Kasper, Markus and Arsenault, Robin and Jakob, Gerd and K{\"a}ufl, Ulli and Leveratto, Serban and Maire, A-L and Pantin, Eric and others},
  journal={Nature Communications},
  volume={12},
  number={1},
  pages={922},
  year={2021},
  publisher={Nature Publishing Group UK London}
}

@article{torres2021canaricam,
  title={CanariCam Mid-infrared Drift Scanning: Improved Sensitivity and Spatial Resolution},
  author={Torres-Quijano, Am{\'\i}lcar R and Packham, Christopher and Acosta, Sergio Fernandez},
  journal={\pasp},
  volume={133},
  number={1029},
  pages={114501},
  year={2021},
  publisher={IOP Publishing}
}

@article{ohsawa2018slow,
  title={“Slow-scanning” in Ground-based Mid-infrared Observations},
  author={Ohsawa, Ryou and Sako, Shigeyuki and Miyata, Takashi and Kamizuka, Takafumi and Okada, Kazushi and Mori, Kiyoshi and Uchiyama, Masahito S and Yamaguchi, Junpei and Fujiyoshi, Takuya and Morii, Mikio and others},
  journal={\apj},
  volume={857},
  number={1},
  pages={37},
  year={2018},
  publisher={IOP Publishing}
}

@article{bradley2016photutils,
  title={Photutils: Photometry tools},
  author={Bradley, Larry and Sipocz, Brigitta and Robitaille, Thomas and Tollerud, Erik and Deil, Christoph and Vin{\'\i}cius, Z{\`e} and Barbary, Kyle and G{\"u}nther, Hans Moritz and Bostroem, Azalee and Droettboom, Michael and others},
  journal={Astrophysics Source Code Library},
  pages={ascl--1609},
  year={2016}
}

@article{sauter2024detection,
  title={Detection Limits of Thermal-infrared Observations with Adaptive Optics. I. Observational Data},
  author={Sauter, JR and Brandner, W and Heidt, J and Cantalloube, F},
  journal={\pasp},
  volume={136},
  number={9},
  pages={095001},
  year={2024},
  publisher={IOP Publishing}
}

@book{lenzen2006statistical,
  title={Statistical regularization and denoising},
  author={Lenzen, Frank},
  year={2006},
  publisher={na}
}

@article{gonzalez2016low,
  title={Low-rank plus sparse decomposition for exoplanet detection in direct-imaging ADI sequences-The LLSG algorithm},
  author={Gomez Gonzalez, CA and Absil, Olivier and Absil, P-A and Van Droogenbroeck, Marc and Mawet, Dimitri and Surdej, Jean},
  journal={\aap},
  volume={589},
  pages={A54},
  year={2016},
  publisher={EDP Sciences}
}

@article{pueyo2016detection,
  title={Detection and characterization of exoplanets using projections on karhunen--loeve eigenimages: Forward modeling},
  author={Pueyo, Laurent},
  journal={\apj},
  volume={824},
  number={2},
  pages={117},
  year={2016},
  publisher={IOP Publishing}
}

\begin{appendix}
\onecolumn
\section{Time-integrated SOFIA data for the F197 filter}
\begin{figure}[!h]
\centering
  \includegraphics[width=.9\linewidth]{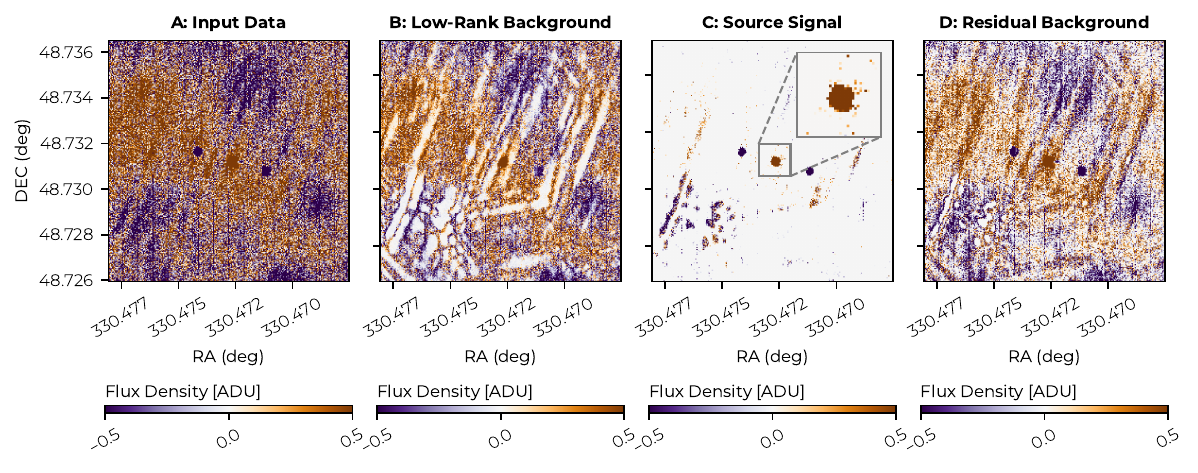}
  \caption{Time-integrated data from a single F197 spectral filter {of SOFIA data}, illustrating the separation of low-rank background, sparse source signal, and photon shot noise achieved by LORABEL. This decomposition allows for clearer identification of the source signal even in the presence of significant background fluctuations. Some residuals still appear (mainly) in the bottom-left of the $C$ image, which is undesirable. In this dataset, it is caused by a lack of time frames and the variability of this part of the background structure (i.e. only appearing in a few frames). Changing the parameters of our method, in this case, could not prevent this.}
  \label{fig:plot5}
\end{figure}
% \newpage
\section{Background subtraction comparison for SOFIA data}
\begin{figure}[!h]
\centering
  \includegraphics[width=.9\linewidth]{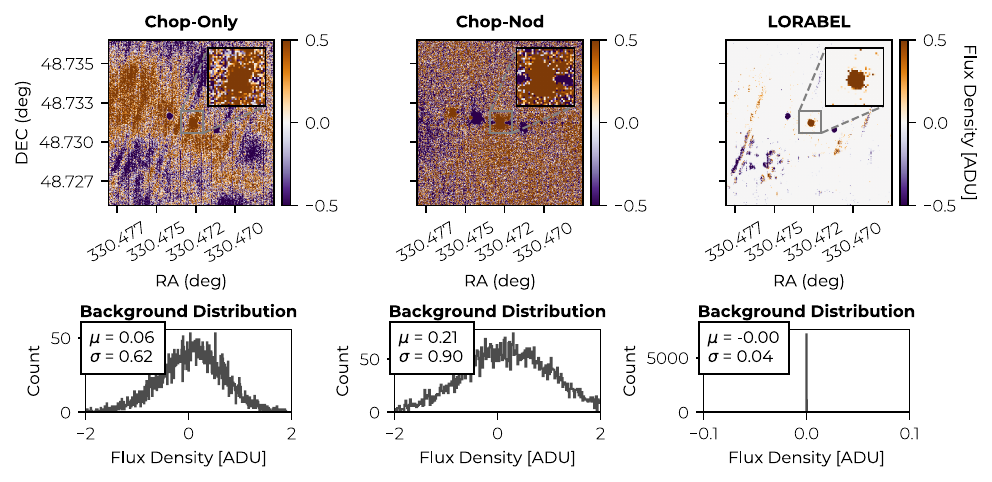}
  \caption{Comparison of background subtraction methods for SOFIA mid-infrared observations. Top panels: Time-integrated images showing the sky background residuals for the Chop-Only (left), Chop-Nod (center), and LORABEL (right). Insets provide zoomed views around the source. Bottom panels: Flux density distributions measured in an annular region of $117$ pixels surrounding the source (with the source masked). The mean ($\mu$) and standard deviation ($\sigma$) of the background residuals are indicated. LORABEL exhibits superior background subtraction, with a notably reduced spread in background flux density compared to traditional techniques. Note that the chop-nod method shows increased photon shot noise due to the combination of two measurements (increased by a factor of $\sqrt{2}$, i.e, $\sqrt{2}\times 0.62 \approx 0.90$). Additionally, small misalignments in the SOFIA data may contribute to incomplete background reduction. Note that the Chop-Only here corresponds to the same ``A: Input Data'' as in Fig. \ref{fig:plot5} (due to scale differences the colours might not appear equal).}
  \label{fig:plot7}
\end{figure}
\end{appendix}
\end{document}